\documentclass[aps,pre,onecolumn,showpacs,a4paper]{revtex4}

\usepackage{graphicx} 
\usepackage{amsmath}

\newcommand\eref{Eq.~\eqref}%
\newcommand\Eref{Equation~\eqref}%
\newcommand{\fref}[1]{Fig.~\ref{#1}} %
\newcommand{\rme}{\mathrm{e}} %
\newcommand{\rmd}{\mathrm{d}} %
\providecommand{\abs}[1]{\lvert#1\rvert}
\providecommand{\Or}{\ensuremath{O}}

\begin{document}

\title{Statistical properties of the final state in one-dimensional
  ballistic aggregation}

\author{Satya N. Majumdar}

\affiliation{Laboratoire de Physique Th\'eorique et de Mod\`eles
Statistiques (UMR 8626 du CNRS), Universit\'e Paris-Sud, B\^atiment 100
91405 Orsay Cedex, France}

\author{Kirone Mallick} 

\affiliation{Institut de Physique Th\'eorique Centre d'\'Etudes de Saclay,
91191 Gif-sur-Yvette Cedex, France}

\author{Sanjib Sabhapandit}

\affiliation{Raman Research Institute, Bangalore 560080, India}

\date{\today}
\pacs{68.43.Jk, 02.50.-r, 05.40.-a, 47.70.Nd}

\begin{abstract}
     We investigate the long time behaviour of the one-dimensional 
     ballistic aggregation model that  represents  a  sticky gas
     of $N$  particles  with  random initial  positions and velocities, 
      moving deterministically,  and  forming  aggregates when they collide.
      We obtain  a closed  formula for the stationary measure of the system
      which allows us to analyze  some remarkable  features  of 
      the final  `fan' state. In particular, we identify 
      universal properties which are independent of  
      the initial position and velocity   distributions  of the particles.
      We study  cluster   distributions  and 
      derive  exact  results for  extreme value statistics 
      (because of correlations  these distributions do not  belong
        to the    Gumbel-Fr\'echet-Weibull
        universality classes). We also derive the energy distribution
       in the final state. This model    
       generates dynamically  many different scales and can be viewed
      as  one of the  simplest exactly solvable model of $N$-body dissipative
      dynamics. 

\end{abstract}

\maketitle

\section{Introduction}

The ballistic aggregation model  is one of the simplest interacting
particles processes of non-equilibrium statistical mechanics and as such
has attracted a lot of attention in the past decades \cite{carnevale}. 
This model represents  a gas of $N$ unit-mass  particles 
 forming clusters through perfectly 
inelastic adhesive collisions. The motion of a particle between two collisions
is deterministic and free (i.e. ballistic) and  the total
mass and momentum  in a collision  are conserved whereas the  kinetic energy is dissipated.
The stochasticity in this model is due only to the initial configuration which
consists of single particles randomly located  with  uncorrelated random  velocities
drawn  from a continuous distribution.  This dissipative
system, usually referred to as ballistic aggregation or sticky gas, appears
as a minimal model of cluster formation and provides a relevant statistical
description of the merger of coherent structures such as vortices, thermal
plumes, flowing granular media~\cite{granular gas} or the accumulation of
cosmic dust into planetoids.  The ballistic aggregation model also plays a
role in the study of the large-scale structure of the universe because the
motion of self-gravitating matter in the expanding universe is similar to
that of matter moving solely by inertia~\cite{Shandarin,martin}.  Another
noteworthy feature of this  model stems from its
connection with the Burgers equation, an important toy model in the study
of turbulence: it can be shown that at very high Reynolds number, the
solution of the Burgers equation in the long time limit consists of a
series of shocks which follow exactly the dynamics of the ballistic
aggregation model~\cite{burgers, kida, frisch}.

 Ballistic aggregation  exhibits at long time $t$ a self-similar
coarsening behaviour first studied   by
Carnevale, Pomeau and Young~\cite{carnevale}. 
 In one dimension,   the
coarsening regime  occurs  for intermediate times satisfying $t \sim
N^{3/2}$,   $N$  being  the total mass or equivalently the
total number of initial particles in the system. For
longer times, and in the absence of any boundary, the system ultimately
reaches a state where no more collisions are possible: the particles are
grouped into clusters of different sizes (or masses) with velocities
increasing from left to right. This final ordered state will be called the
``fan'' state. Once this state is reached, there is no further loss of
energy and the system becomes stationary.

The aim of this paper is to investigate the properties of this fan
state. We find an exact analytic formula for the joint probability
distribution of the sizes and velocities of the clusters by mapping the
final state of the ballistic aggregation model to the convex minorant of a
one-dimensional random walk. From this invariant measure that fully
characterizes the set of all possible fan states, various statistical
properties of the  model in the long time limit will
be derived.  In particular, we shall retrieve the
  known fact \cite{shida,sibuya, hyuga}
that the probability of obtaining a fan state with exactly $k$ clusters is
a purely combinatorial factor (which is universal in the sense that it does
not depend upon the initial velocity  distribution of the 
particles). We shall prove that the cluster distribution in the fan state is
identical to the statistics of cycles in a random permutation of $N$
objects \cite{goncharov, shepp}, both problems having a common underlying
combinatorial structure related to the convex minorant construction
\cite{spitzer,steele}.  This will allow  us to characterize the different
scales that are dynamically generated in the system:  in the large $N$ limit,
 the size of a typical cluster is
of order $N/\ln N$, the largest cluster contains a finite fraction of the
total mass $N$ and hence grows linearly with $N$,
the smallest cluster scales as $\ln N$, and 
the rightmost
cluster, which has the leading velocity, has a size of order $\sqrt{N}$.
We shall also derive the (non-universal) joint distribution of  mass and
velocity of this `leader' which is very different from the usual
extreme-value distributions obtained for uncorrelated random variables.
Finally, we shall show that the energy remaining in the system after all
collisions have ended is of the order $\ln N$ and  shall calculate the
distribution of the scaling variable $E/\ln N$.

The paper is organized as follows. In Section-II we define the ballistic
aggregation model  and focus on its final fan state. In particular,
 we show how the statistical properties in the fan state are
related to those of the convex minorant of an associated 
one dimensional random walk problem. This mapping allows us to derive
explicitly the joint distribution of the cluster sizes and their velocities
in the fan state which is one of the  key results of this
paper. In Section-III we show how this joint distribution can be used to
derive the cluster size distribution in the fan state which turns
out to be universal (with respect to the initial velocity distribution).
This universal property is shown to be related to the statistics of cycles
lengths in a random permutation problem. In Section-IV, we further exploit
the joint distribution to compute exact asymptotic results for certain
extreme variables in the fan state such as the size and the velocity
distribution of the rightmost (`leader') cluster. In Section-V, we derive 
the exact asymptotic distribution of energy in the fan state.  Finally we
conclude in Section-VI with a summary and some open problems.

\section{Statistical Description of the Stationary State}

\subsection{Definition of the Model}

   The ballistic aggregation problem in one dimension is an elementary
 interactive particles process, the dynamics of which is defined as follows.
 At the initial time  $t=0$, we start with  $N$ particles randomly
 located on an infinite line. We  label the particles as $i=1,2,\dotsc,N$
 according to  the order of their initial positions from left to
  right (see Fig.~1 where we have taken $N=6$).  We denote by $v_i$ and $m_i$, 
 respectively,  the initial velocity and the mass of the $i$-th particle.
 For $t > 0$, each particle moves ballistically at constant speed.
 When two particles meet they coalesce and  form  a  single new particle 
  whose mass is the sum of the initial  masses and whose momentum
 is the sum of the initial momenta. The only randomness in the 
model lies 
in the
 initial conditions, i.e. the initial positions and velocities 
 of each particle. We shall assume   that the initial
 velocities  $\{v_i\}$'s are uncorrelated random variables, drawn from a
common continuous
  probability density function (PDF) $\phi(v)$.  For simplicity, we
assume that  all the initial masses are the same  and take them  to  be unity.
  Once the initial conditions are given, the evolution
 of a given realization is  deterministic and fully determined by the mass
 and momentum conservation laws. The shocks being totally inelastic,
   energy is dissipated at each collision. 

At late times $t\gg O(1)$ (but before the system feels the finiteness
of the mass $N$), 
the ballistic aggregation model exhibits a coarsening regime
where typical cluster size grows with time as a power law, as
studied first by
Carnevale, Pomeau and Young~\cite{carnevale} using scaling arguments and
numerical simulations. For example, in this regime, 
 the average size (or mass) of a particle-cluster grows as
$t^{2D/(2+D)}$ and  its velocity decreases as $t^{-D/(2+D)}$,  $D$ being 
 the spatial dimension. The global mass of the system being conserved,
the total energy of the system decays as $t^{-2D/(2+D)}$. 
These exponents derived in ~ \cite{carnevale} turn out to be correct in one dimension
where the problem is exactly solvable~\cite{martin,frachebourg,fracheb2}. 
This, however, turns out to be a bit fortuitous.
In higher dimensions,
the exponents predicted in ~\cite{carnevale} turn out to be incorrect
due to strong correlations between the velocities of the colliding clusters
at late times~\cite{PaulK}.    
In one dimension,   it is possible to calculate exactly (in  one dimension)
  the mass distribution
 of the clusters \cite{martin, frachebourg, fracheb2}
 in  the scaling limit when $N\to \infty$ ($N$  being  the total mass of the system)
  and $ t \to \infty$, keeping the ratio
 $t/N^{3/2}$ finite   \cite{martin, frachebourg, fracheb2}.
  However,  these  scaling results for  the
coarsening regime  are valid only for intermediate times satisfying $t \sim
N^{3/2}$. When $t \gg N^{3/2}$, the system evolves into
 a stationary state in which no  more collisions can occur:
 the particles are grouped in $k$ disjoint
 clusters of different masses, where $k$ itself is a stochastic variable. 
Each  cluster  moves  
at a constant
velocity,  and the speed of a  given cluster is larger than that of its
left neighbour (if any) and less than that of its right  neighbour (if any). 
In this ultimate state the clusters keep  on moving farther apart, i.e., they fan out 
from
each other with increasing time, thus justifying the name ``fan" state (see Fig.~1). 
Once this fan state is reached, there is no further collision and hence
no further dissipation of energy.
 
\begin{figure}
\includegraphics[width=3.375in]{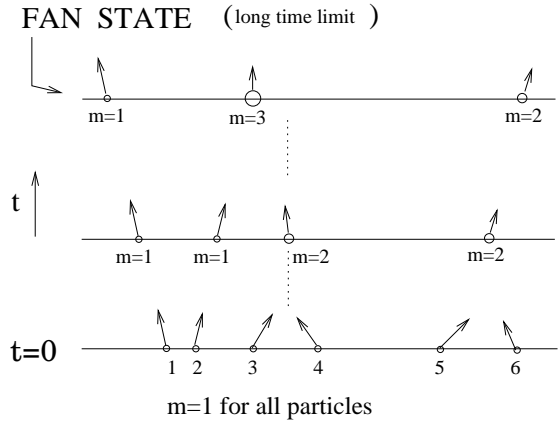}
\caption{\label{fan} An example of a configuration of $6$ particles with
identical initial mass $m_i=1$ and  varying initial velocities,
evolving in the long time limit into the fan state: the velocities are now 
increasing from left to right and there are no more  colllisions.}
\end{figure}

  In this work, we shall focus on the statistical description
  of the fan state of the random aggregation process. We shall
  show that  properties of the  fan state related  to the sizes 
 of the clusters (regardless of their respective ordering
 and their velocities)  are universal as they do not
 depend on the PDF  $\phi(v)$ from which the initial velocities
 of the particles are drawn. The cluster statistics can in fact 
 be mapped  to the  cycle length distribution in  random permutations, 
 which are  fundamental combinatorial objects. From this observation
the distribution of the cluster masses  and in particular the typical
 mass of the largest and the smallest cluster can readily be calculated.
 When the distribution of the velocities of the clusters is taken into
 account, universality with respect to $\phi(v)$ is lost.
 However, in the large $N$ limit, we will see that some universality
is restored, thanks to the central limit theorem, in the size distribution of the rightmost 
cluster (the leader)
provided the second moment $\sigma^2=\int_{-\infty}^{\infty} v^2\,\phi(v)\, dv$ is finite.  
Finally, we are also able to calculate explicitly the energy distribution in the fan state, 
but
only for the special case when $\phi(v)$ is Gaussian.

\subsection{The Fan State as a  Convex  Minorant} 

The fan state of the random aggregation process can be determined
geometrically from the initial conditions~\cite{sibuya}. 
We now  explain this  graphical interpretation  of the fan state as a convex
minorant of a random walk.    Recall that we  label the particles
 as $i=1,2,\dotsc,N$ according to the  order of their initial positions from left to
  right (see Fig.~1) and that  we  denote by $v_i$ 
 respectively the initial velocity  of the $i$-th particle (all the
 initial masses being unity).  For any
realization of the initial state  $\{v_1,v_2,\ldots,v_N\}$, the final state
 is uniquely determined i.e. the number of final clusters, 
their masses and their  velocities are unique. 

 The initial state  $\{v_1,v_2,\ldots,v_N\}$ of the system is represented
 by a broken graph $(P_0, P_1, \ldots, P_N)$ (see Fig.~2) such that: 

 (i) $P_0 =(0,0)$ is the origin;

 (ii) the coordinates of $P_i$ are given recursively by
  $P_i = P_{i-1} + (1,v_i)$ for $1\le i \le N$.

 In other words, the speed $v_i$  of the $i$-th particle is represented
 by the {\it slope} of the line $(P_{i-1}P_i)$. We also emphasize  that
 the horizontal  coordinate  of  $P_i$  does not correspond to the
 actual  position of the $i$-th particle but only on its label.

\begin{figure}
  \includegraphics[width=3.50in]{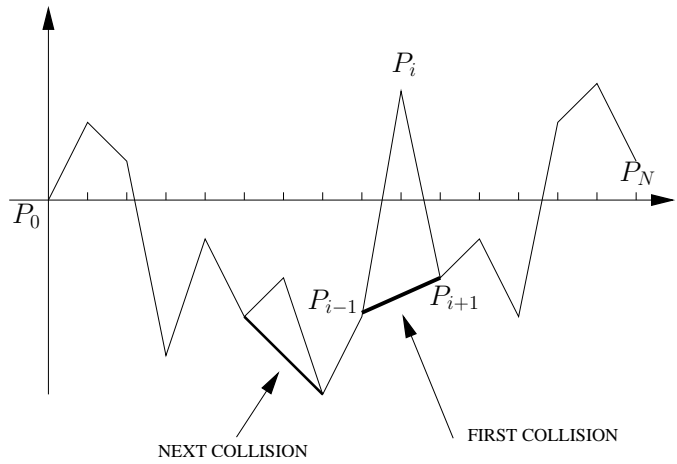}
   \caption{\label{minorant2} Initial configuration is represented by a
    graph  (thin line) joining  $P_0$ to $P_N$
   (we have taken $m_j=1\; \forall j$). This graph can be interpreted as 
   a random walk in the cumulative momentum and cumulative mass space.
  After the first  collision the particles  $i$ and $i+1$ 
     aggregate into one cluster:  in the associated  graph, the angle
  $(P_{i-1}P_{i}P_{i+1})$  is replaced by the 
  segment $(P_{i-1}P_{i+1})$.}
\end{figure}

\begin{figure}
  \includegraphics[width=3.5in]{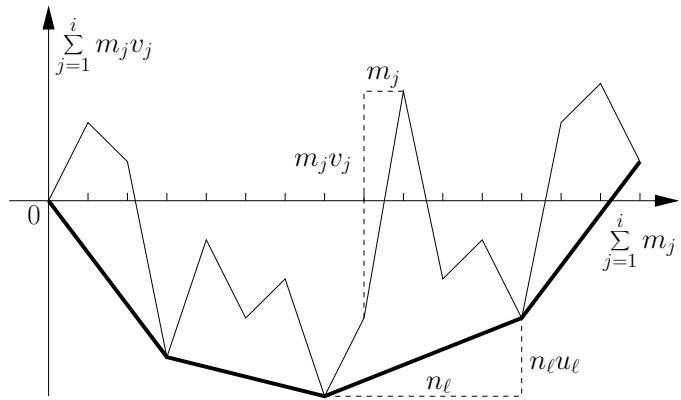}
  \caption{\label{minorant}  We draw the convex minorant
 of the graph representing   the same   initial configuration
 as in Fig.~\ref{minorant2}. This convex minorant
 represents the final `fan'  state in which no more collisions can occur.
 Each straight segment of the  convex minorant corresponds to a cluster
 of $n_l$ particles with momentum $n_l u_l$.}
\end{figure}

 More generally, if we had started with particles having different masses
  $\{m_1,m_2,\ldots,m_N\}$  
  the coordinates of $P_i$ would be  
  $P_i = P_{i-1} + m_i(1,v_i)$~\cite{sibuya}. In other words,
 the initial state is drawn as a  random walk in the cumulative momentum and
cumulative mass space as shown in  Figs.~\ref{minorant2} and \ref{minorant}. 

 Suppose that the first collision
 occurs  between  particles $i$ and $(i+1)$
 with velocities $v_i$ and $v_{i+1}$ respectively. These two particles can
 aggregate if  $v_i > v_{i+1}$, which means that 
 the slope of the  $(P_{i-1}P_i)$ is larger than the 
 slope of the  $(P_{i}P_{i+1})$. Equivalently, this means that
 $P_{i}$ is located above the segment $(P_{i-1}P_{i+1})$ i.e.,  
 locally,  the graph  $(P_{i-1}P_{i}P_{i+1})$ has a negative  curvature.
 After this collision, a cluster is formed 
 with  mass 2  and velocity $(v_i + v_{i+1})/2$. This cluster
 is represented by the vector  $(P_{i-1}P_{i+1})$
  (see Fig.~\ref{minorant2}) 
 with coordinates $(2, v_i + v_{i+1}).$  We note that
 the slope of this vector again represents  the velocity of the cluster.
 We now have
 a system  of $N-1$ `particles', with $N-2$ particles of mass 1 
 and one  particle of mass 2. The state of this system is  
 represented by  a broken line   in which
 the angle   $(P_{i-1}P_{i}P_{i+1})$   with downward curvature 
 is replaced by its base  (see~\fref{minorant2}). 
 Similarly, the  next  collision is also  represented graphically
 by replacing another  angle with downward curvature by its base
 (see Fig.~\ref{minorant2}) . 
 This process will continue  iteratively and the particles will aggregate
 forming clusters till all  angles with downward curvature have
 been eliminated, i.e. all collisions have occurred. 
 It follows that for any given initial state, the
final state is uniquely given by the convex minorant of the corresponding
random walk. Each line segment of the convex minorant represents a cluster
in the fan state, ---the horizontal and the vertical components of the
segment give respectively   the mass and the momentum of the cluster, i.e.,
the slope gives the velocity of the cluster. The geometry of the convex
minorant (including the number of segments which represents the number of
final clusters), however, differs from one realization of initial
conditions to another. Various statistical properties of the ballistic
aggregation in the fan state can be found from the statistical properties
of the underlying convex minorant.

 \subsection{The Invariant  Measure in the Fan  State}

We now derive the joint probability distribution which describes the fan
state.  Let $p_N(k;\{n_i,u_i\})$ denote the joint PDF of having $k$
clusters in the fan state  of masses and velocities $n_i$ and $u_i$
respectively, with  $i=1,2,\dotsc,k$.  Note that the various segments of the
convex minorant (see \fref{minorant}) are correlated as their slopes must
be increasingly ordered from left to right, i.e.,
$u_1<u_2<\dotsb<u_k$. Moreover, the total mass is conserved, $\sum_{i=1}^k
n_i=N$. Once these two constraints are specified, the momentum of each
final cluster is essentially determined by the conservation law:
\begin{equation}
  n_\ell u_\ell=
       \sum_{j = N_{\ell-1}+1}^{N_\ell} m_j v_j
\end{equation}
 for the $\ell$-th cluster with
$\ell=1,2,\dotsc,k$.  We have defined here the cumulated masses
 $N_k = n_1 + \ldots + n_k$.
 The sum in the above  equation is thus  restricted  to  the
initial velocities of those particles which form the $\ell$-th cluster.
 We also  
recall that  $m_j=1\; \forall j$.
  However, most importantly, to form a specific
(say $\ell$-th) final cluster (i.e., a line segment in the underlying
convex minorant) by aggregation of $n_\ell$ consecutive initial particles, 
 the portion of the underlying random walk of $n_\ell$ steps
must be such that it originates from one end of the line segment of convex
minorant and finishes at its other end, while staying above the segment in
the intermediate steps.  Now,  consider a random walk of $n_\ell$
steps that starts from one end of a line segment and finishes at its  other
end, but is  otherwise  allowed to cross the segment in the
intermediate steps.  For any realization of such  a walk, given that there is
one unique minimum with respect to the segment, if we consider all the
$n_\ell$ cyclic permutations of the steps, then out of $n_\ell$ different
realizations of the walks there is  one and only one  arrangement where
in the intermediate steps the walk stays above the
  segment.  This property  is known as Raney's lemma~\cite{feller,Knuth}.
  (Here,  the uniqueness of the arrangement is  guaranteed  by the fact that the  PDF  
$\phi(v)$ is  continuous).

Thus, the probability that a random walk of $n_\ell$ steps that starts at
one end of a line segment and finishes at its other end while staying above
the segment in-between equals $1/n_\ell$.  Note that this probability is
independent of the slope $u_\ell$ of the segment as well as the
PDF $\phi(v)$ of the jump-lengths.  Finally, gathering all the above inputs
together we write
\begin{align}
p_N(k;\{n_i,u_i\})&=\delta\left(N-\sum_{i=1}^k n_i
  \right)\,\prod_{i=1}^{k-1}\theta(u_{i+1}-u_i)\notag\\
&\times\prod_{\ell=1}^k \frac{1}{n_\ell}\left\langle\delta\left(
 u_\ell-\frac{1}{n_\ell}\sum_{j=N_{\ell-1}+1}^{N_\ell}  v_j
\right)\right\rangle,
\label{jointpdf}
\end{align}
where the angle brackets $\langle \dotsb\rangle$ denote the averaging over
the set of initial velocities $\{v_i\}$.  This key  result
  provides a full statistical  description of the clusters sizes
 and velocities   in the fan state.
 From this joint probability distribution,  all the statistical properties
  of the final state can be derived.

\section{Universal Statistical Properties of Clusters in the Fan State}

 In this section, we compute the distribution of the number and 
 sizes of the clusters in the fan state:
we will  show 
 that this distribution,  obtained by integrating out the 
 velocities of the clusters  from  the
  joint probability distribution given in Eq.~(\ref{jointpdf}), 
 is universal with respect to the initial continuous PDF $\phi(v)$.

\subsection{Statistics of the Total Number of Clusters in the Fan State}

  We first  calculate  $p_N(k)$, 
  the probability that the final state contains $k$ distinct clusters.
  The value of  $p_N(k)$ is obtained by summing Eq.~(\ref{jointpdf}) 
 over all  $n_i>0$'s  and integrating it over  $u_i$'s, keeping $k$ fixed.
 Thus, 
\begin{align}
p_N(k) = \sum_{\{n_i\}} \int \prod_{i=1}^k du_i \, 
 \delta\left(N-\sum_{i=1}^k n_i
  \right)\,\prod_{i=1}^{k-1}\theta(u_{i+1}-u_i)
  \prod_{\ell=1}^k \frac{{\mathcal P}(n_\ell, u_\ell) }{n_\ell} \, .
\label{Nbcluster1}
\end{align}
 In this equation,  we have used 
 \begin{align}
{\mathcal P}(n_\ell, u_\ell) = \left\langle\delta\left(
u_\ell-\frac{1}{n_\ell}\sum_{j=N_{\ell-1}+1}^{N_\ell}  v_j 
\right)\right\rangle =   \int   \, \delta\left( u_\ell-
\frac{1}{n_\ell}\sum_{j=N_{\ell-1}+1}^{N_\ell}  v_j \right)
 \prod_{j = N_{\ell-1}+1}^{N_\ell}  \phi(v_j) dv_j \, ,
\label{def:mathcalP}
\end{align}
 where $\phi$ is the distribution of the initial velocities.
 We note that  the function ${\mathcal P}(n, u)$ is normalized:
\begin{align}
 \int_{-\infty}^{+\infty} du {\mathcal P}(n, u) = 1 \, .
\label{NormP(n,u)}
\end{align}

 To disentangle the constraints in Eq.~(\ref{Nbcluster1}), it is useful 
 to consider the following  generating function:
\begin{align}
 \sum_N p_N(k) z^N = \sum_{\{n_i\}} \frac{z^{n_i}}{n_i}
 \int \prod_{i=1}^k du_i  \, \prod_{i=1}^{k-1} \theta(u_{i+1}-u_i)  \, \, 
  {\mathcal P}(n_i, u_i)  = 
 \int \prod_{i=1}^k du_i  \, \prod_{i=1}^{k-1} \theta(u_{i+1}-u_i) \,
    \rho_z(u_i)
 \label{FonctGen1}
\end{align}
 with 
\begin{align}
\rho_z(u_i) =  \sum_{\{n_i\}} \frac{z^{n_i}}{n_i}  {\mathcal P}(n_i, u_i) \, .
\label{defrho}
\end{align}
 The integral on the right hand side (r.h.s) of Eq.~(\ref{FonctGen1}) can be
 evaluated recursively and is found to be 
\begin{align}
\int_{-\infty}^{+\infty} du_k  \rho_z(u_k)
 \int_{-\infty}^{u_k}  du_{k-1} \rho_z(u_{k-1}) \ldots 
   \int_{-\infty}^{u_2}  du_{1} \rho_z(u_{1}) = \frac{1}{k!} 
 \Big( \int_{-\infty}^{+\infty}  du_k  \rho_z(u_k) \Big)^k = 
 \frac{1}{k!}  \left( \sum_{\{n_i\}} \frac{z^{n_i}}{n_i} \right)^k \, ,
\label{thetaprod}
\end{align}
where the last equality is obtained  by using the 
 definition of $\rho_z(u)$ in (\ref{defrho})
 and the normalization property in (\ref{NormP(n,u)}).
 Thus, for any given value of $k>0$, we have 
\begin{align}
 \sum_N p_N(k) z^N = \frac{1}{k!}\Bigl[-\ln (1-z)\Bigr]^k \, . 
\label{FGen1}
\end{align}
 After extracting the coefficient of $z^N$ on the r.h.s of this equation,
 we conclude that
\begin{align}
p_N(k)
 &=\frac{1}{k!} \sum_{\{n_i>1\}}^\infty \left[\prod_{i=1}^k
   n_i\right]^{-1} \delta\left(N-\sum_{i=1}^k n_i \right) \, .
\end{align}
  This formula can be expressed  in terms of 
 some  classical combinatorial numbers as follows. 
 It is convenient to introduce a second auxiliary variable $t$ and
 to calculate the generating function of Eq.~(\ref{FGen1}) with respect to 
 $t$:
 \begin{align}
\sum_k \sum_N p_N(k) z^N t^k = \sum_k
 \frac{1}{k!}\Bigl[-\ln (1-z)\Bigr]^k t^k =  (1-z)^{-t} 
 = 1 + tz + \frac{t(t+1)}{2!} z^2 +  \frac{t(t+1)(t+2)}{3!} z^3 + \ldots
\label{FGen2}
\end{align}
  Identifying the coefficients of $z^N$ on both sides of this equation
 we obtain the identity:
 \begin{align}
 \sum_k   p_N(k)  t^k = \frac{t(t+1)(t+2)\ldots(t+N-1)}{N!} \, .
\label{FGen3}
\end{align}
 Using the classical formula
\begin{align}
 t(t+1)(t+2)\ldots(t+N-1)= \sum_{k=1}^N S_1(N,k)\, t^k \, ,
 \label{stirling gen 2}
\end{align}
 that defines the   unsigned Stirling numbers 
 of the first kind \cite{riordan}, 
$S_1(N,k)$, we conclude that the probability of finding
 $k$ clusters in the fan state for a system starting with $N$
 distinct unit-mass particles is given by
\begin{align}
p_N(k) = 
&=\frac{S_1(N,k)}{N!}, \quad\text{with}~ k=1,2,\dotsc,N \, .
\label{pk}
\end{align}
 We  thus recover by a different method  the result of
Ref.~\cite{sibuya}. 

 From \eref{stirling gen 2}, we observe that
$\sum_{k=1}^N S_1(N,k)=N!$. Thus, $p_N(k)$, and consequently the joint PDF
given by \eref{jointpdf} are both clearly normalized as it must.
 Furthermore,  using Eq.~(\ref{FGen3}), one deduces  that
 the mean number of clusters is   given by 
 $1 + \frac{1}{2} + \ldots + \frac{1}{N}$.

 In fact, it is well  known   \cite{riordan} that the   unsigned
Stirling numbers 
$S_1(N,k)$ enumerate  the number of permutations of $N$ elements with 
$k$ disjoint cycles exactly:  thus, $p_N(k)$ is identical to the probability
distribution of number of cycles in random permutations with uniform
measure.  Many results have been derived concerning the statistics
 of random permutations. For example,  it has  been shown
 in the context of permutation cycles~\cite{shepp}
that for large $N$, $p_N(k)$ has a Gaussian form around its mean $\langle
k\rangle \sim \ln N$, with a variance which is also of order $\ln N$. 
Therefore, the {\em typical} mass of a  cluster is   of
  order   $N/\ln N$.  Using
this together with the $t^{2/3}$ asymptotic growth of cluster sizes before
the  fan  state is reached, the characteristic time to reach this 
state is  estimated to be  $t_c \sim (N/\ln N)^{3/2}$.


\subsection{Distribution of Clusters Sizes including the Biggest and the Smallest}

    We now consider a more refined observable.  It may be tempting
 to calculate the  cluster size distribution i.e. to
 integrate out the velocities from Eq.~(\ref{jointpdf}) and determine
  the marginal PDF $p_N(k;\{n_i\}) $. Unfortunately, this marginal 
 distribution is not universal:  it depends on the choice
 of the initial PDF $\phi(v)$ (this fact can be verified by working out
  explicit  examples).  However, if one  considers 
  the statistics of the clusters
 according to their sizes regardless of their  spatial ordering
  then the result is again $\phi(v)$-independent.

 More precisely, let $c_j$ for $j=1,2,\dotsc,N$
  denote the number of clusters of
 size $j$ in the fan state. By mass conservation,
  we  have  $\sum_j j c_j = N$.
 Then,  it can be shown
  that the joint probability, $\text{Pr}\{c_1,c_2,\dotsc,c_N\}$
 is given by 
\begin{align}
\text{Pr}\{c_1,c_2,\dotsc,c_N
 \}=\delta\bigl(N-\sum_{j=1}^N j c_j\bigr)\prod_{j=1}^N \frac{1}{j^{c_j} c_j!}
  \, .  \label{Prclustersizes}
\end{align}
 The derivation  of this formula  from the invariant measure
 in the fan state, Eq.~(\ref{jointpdf}),  is given in the Appendix A.
  This expression is identical to the distribution of 
  cycles in random permutations where $c_j$'s are to be identified with the
  number of cycles of length $j$ \cite{riordan}.

The generating function associated with  this probability distribution
 is defined as follows:
\begin{align}
 G_N(t_1, \ldots, t_N) =
  \sum_{{c_1,\ldots,c_N}} \text{Pr}\{c_1,c_2,\dotsc,c_N\}
 t_1^{c_1} \ldots t_N^{c_N} \, .
\end{align}
 This  function  $G_N$ can itself be embeded in a 'grand-canonical'
  generating function given by
\begin{align}
 \Gamma(z, \{t_k\}) = \sum_{N}G_N(t_1, \ldots, t_N) z^N
              = \exp\Big( \sum_{k=1}^\infty \frac{z^k t_k}{k}  \Big) \, ,
 \label{GrandCan}
\end{align}
 where the last equality is obtained by using Eq.~(\ref{Prclustersizes}).
 This expression embodies all the necessary  information and will be 
 useful for further calculations.

   For example, we can study
  the extreme cluster sizes.  Let 
$Q_N(L)$ be the
probability that the largest cluster size (mass) $n_{\max}\le L$ and 
$R_N(S)$  the probability
that the smallest cluster size $n_{\min} \ge S$. It follows
 from these definitions that 
the probability that the largest cluster is of size $L$ is given by
\begin{align}
\text{Pr} \left( n_{\max} = L \right) = Q_N(L) - Q_N(L-1) \, , 
\end{align}
and similarly 
\begin{align}
\text{Pr} \left( n_{\min} = S \right) = R_N(S) - Q_N(S+1) \, .
\end{align}
 Using    Eq.~(\ref{Prclustersizes}) we find   
\begin{align}
\label{Q_N}
Q_N(L) =  \sum_{c_1,\ldots,c_L} \text{Pr}\{c_1,c_2,\dotsc,c_L,0,\dotsc,0\}
 =      G_N(t_1=1, \ldots,t_L=1,0,\ldots,0)           \, , 
\end{align}
and similarly 
\begin{align}
  R_N(S) =  \sum_{c_{S},\ldots,c_N} \text{Pr}\{0,\dotsc,0,c_{S},\ldots,c_N\}
        = G_N(0, \ldots,0, t_S=1, \ldots, t_N=1) \, .
\end{align}
 Using the grand-canonical generating function defined in Eq.~(\ref{GrandCan}), 
 we readily obtain
\begin{align}
  Q_L(z) =
 \sum_N Q_N(L) z^N = \exp\left(\sum_{n=1}^L \frac{z^n}{n} \right) \,\,
\hbox{ and } \,\,\, R_L(z)
\sum_N R_N(S) z^N =\exp\left(\sum_{n=S}^\infty \frac{z^n}{n} \right).
\label{GFmaxmin}
\end{align}

The mean sizes  of  the largest  and the smallest
clusters in the fan state containing $N$ particles  are given by  
\begin{eqnarray}
\langle n_{\max} \rangle_N &=& \sum_L L[Q_N(L)-Q_N(L-1)] = 
\sum_L [1-Q_N(L)] \\ \hbox{ and }  \,\,\,\,   \langle
n_{\min} \rangle_N &=&  \sum_S S[R_N(S) - R_N(S+1)] =\sum_S R_N(S) \, . 
\end{eqnarray} 
  For large $N$, the asymptotic  behaviour of 
  $\langle n_{\max} \rangle_N$  and $\langle n_{\min} \rangle_N$
can be obtained by analyzing  the associated  generating functions  
 in the $z\rightarrow 1$ limit. For example, using Eqs.~(\ref{GFmaxmin})
   we obtain 
\begin{eqnarray}
 \sum_N  \langle n_{\max} \rangle_N z^N = \sum_L \Big\{  \frac{1}{1-z}
 -  \exp\left(\sum_{n=1}^L \frac{z^n}{n} \right) \Big\}
 = \frac{1}{1-z} \sum_L \Big\{  1 - 
 \exp\left(-\sum_{n=L+1}^\infty \frac{z^n}{n} \right) \Big\} \, . 
\end{eqnarray} 
 When $z \to 1$, the sums on the right-hand sides can be transformed
 into integrals and  the desired asymptotics are  obtained by 
  extracting the leading singularity at $z=1$. Thus,  for large $N$, we obtain 
\begin{equation}
\langle n_{\max} \rangle \sim \lambda N \,\,\, \hbox{ with }  \,\,\,
\lambda=\int_0^\infty
\exp\left(-x-\int_x^\infty  \frac{\rme^{-y}}{y}\, \rmd y\right)
\, \rmd x=0.6243299885\dots
\end{equation}
 This constant is  known as the Golomb-Dickman Constant~\cite{Finch}. Similarly,
 We find, $\langle n_{\min} \rangle\sim\rme^{-{\bf C}}
\ln N$, where ${\bf C}=0.57721566\dots$ is Euler's constant, and
  we  again note that,
$\langle n_{\min}\rangle$ and $\langle n_{\max}\rangle$ are identical to
the mean lengths  of the shortest and the longest cycles, respectively,  of
random permutations  and were already calculated
 in this context \cite{shepp,arratia}. Recently, the same constants 
also appeared in the context of the statistics of longest and shortest
lasting records of independent and identically distributed random variables~\cite{MZ,GL} 


\section{Leader Statistics }

In this section we compute the statistics of the size and the velocity of
the `rightmost cluster' in the fan state. This rightmost cluster will
be referred to as the `leader' since it moves with the highest velocity.
Let $L_N(n,v)$ denote the joint PDF of the leader's size $n$
and velocity $v$ in the fan state. To obtain this PDF, we start from the 
basic microscopic joint PDF of all cluster sizes and their velocities in
Eq. (\ref{jointpdf}) and integrate out the sizes and velocities
of all clusters except the leader. As usual, it is convenient to
consider the generating function which then reads
\begin{equation}
\sum_N L_N(n,v)\, z^N = \frac{z^n}{n} {\mathcal P}(n,v)\sum_{k=1}^{\infty} \int 
\left[\prod_{i=1}^{k-1} 
du_i \rho_z(u_i)\right] \, \theta(v-u_{k-1})\,
\left[\prod_{i=1}^{k-2} \theta(u_{i+1}-u_i)\right] 
\label{lpdf1}
\end{equation}
where $\rho_z(u)$ and ${\mathcal P}(n,v)$ are defined respectively in Eqs. (\ref{defrho})
and (\ref{def:mathcalP}). The $(k-1)$-fold integral over the $u_i$'s can be
recursively evaluated as in Eq. (\ref{thetaprod}) and we get
\begin{equation}
\int
\left[\prod_{i=1}^{k-1}
du_i \rho_z(u_i)\right] \, \theta(v-u_{k-1})\,
\left[\prod_{i=1}^{k-2} \theta(u_{i+1}-u_i)\right]=  \frac{1}{(k-1)!}
 \Big( \int_{-\infty}^{v}\rho_z(u) du \Big)^{k-1}. 
\label{thetaprod2}
\end{equation}
Substituting this in Eq. (\ref{lpdf1}) and summing over $k$ we get a somewhat
compact expression
\begin{equation}
\sum_N L_N(n,v)\, z^N 
= \frac{z^n}{n}\, {\mathcal P}(n,v)\, \exp\left[\int_{-\infty}^v \rho_z(u) du\right].
\label{lpdf2}
\end{equation}
For the subsequent asymptotic analysis, it turns out to be convenient to rewrite
the expression in Eq. (\ref{lpdf2}) in a slightly different form.
Using the definition of $\rho_z(u)$ from Eq. (\ref{defrho}) and the normalization
in Eq. (\ref{NormP(n,u)}), it follows that
\begin{equation}
\int_{-\infty}^v \rho_z(u)du = 
\sum_{k=1}^{\infty}\frac{z^k}{k}\left[1-\int_v^{\infty}{\mathcal P}(k,u)\,du\right]
=-\ln(1-z)-\int_v^{\infty} \rho_z(u) du.
\label{redef}
\end{equation} 
Using this in Eq. (\ref{lpdf2}) gives
\begin{equation}
\sum_N L_N(n,v)\, z^N= \frac{z^n}{n(1-z)}\,{\mathcal P}(n,v)\,
\exp\left[-\int_v^{\infty} \rho_z(u) du\right].
\label{lpdf3}
\end{equation}

From the joint PDF $L_N(n,v)$ of the leader's size and velocity in Eq. (\ref{lpdf3})
one can then obtain the marginals: $C_N(n)= \int_{-\infty}^{\infty} L_N(n,v) dv$ 
for the size PDF
and $D_N(v)= \sum_{n=1}^{\infty} L_N(n,v)$ for the velocity PDF of the leader.
It is evident from Eq. (\ref{lpdf3}) that $L_N(n,v)$, for any finite $N$, depends
explicitly on the initial velocity distribution $\phi(v)$ since both
${\mathcal P}(n,v)$ as well as $\rho_z(u)$ depends on $\phi(v)$. This
is unlike the distribution of the cluster sizes as
derived in Section III which is completely universal, i.e., independent
of $\phi(v)$ for any finite $N$. So, a natural question is: to what
extent the leader size and velocity distributions are universal, i.e.,
independent of the details of $\phi(v)$? We will see below
that the universal property holds for the leader's size distribution, but
only in the large 
$N$ limit and 
for continuous and symmetric $\phi(v)$ with a finite
variance
$\sigma^2= \int_{-\infty}^{\infty} v^2 \phi(v) dv$. In that
case, for large $N$, one can use the central limit theorem and
the limiting marginal size distribution $C_N(n)$ essentially
becomes universal, though the marginal velocity distribution
$D_N(v)$ still remains nonuniversal even in the large $N$ limit.
A natural question is how these results get modified when
$\phi(v)$ is such that its variance $\sigma^2$ is infinite.
A particular example of this class is the Cauchy distribution
for $\phi(v)$. We will show that the Cauchy case is exactly solvable
and one can derive the size and the velocity distribution of 
the leader explicitly.  
In subsections A and B below, we 
consider, for finite $\sigma^2$, the limiting size and the velocity
distribution of the leader respectively.
In subsection C, we derive the explicit distributions for the Cauchy case.

\subsection{ The Limiting Leader Size Distribution}

In this subsection we compute the limiting size distribution $C_N(n)$ of the leader
for large $N$. Integrating Eq. (\ref{lpdf3}) over $v$ and writing $z=e^{-s}$ we get
\begin{equation}
\sum_N C_n(n) e^{-sN}= \frac{e^{-sn}}{n(1-e^{-s})}\,\int_{-\infty}^{\infty} dv {\mathcal 
P}(n,v)\,\exp[-Y(s,v)]
\label{lsd1}
\end{equation}
where 
\begin{equation}
Y(s,v)= \sum_{k=1}^{\infty} \frac{e^{-sk}}{k}\, \int_v^{\infty} du\, {\mathcal
P}(k,u)
\label{lsd2}
\end{equation}
with ${\mathcal P}(k,u)$ defined in Eq. (\ref{def:mathcalP}). Now, for large $N$,
we need to investigate the behavior of the r.h.s of Eq. (\ref{lsd1}) in the limit
$s\to 0$. It is evident that to obtain a sensible limit of the r.h.s
in Eq. (\ref{lsd1}) one needs to consider the scaling limit $s\to 0$ but $n\to \infty$
keeping the product $sn$ finite. In this scaling limit, it is not difficult
to see that the dominant contribution to the sum $Y(s,v)$ in Eq. (\ref{lsd2})
comes from those terms where the product $sk$ is finite in the limit $s\to 0$, i.e.,
terms where $k\sim 1/s$ is large.

Now, for large $k$,
it follows from the definition in Eq. (\ref{def:mathcalP}) that
for a symmetric $\phi(v)$ with zero mean and a finite variance 
$\sigma^2= \int_{-\infty}^{\infty} v^2 \phi(v) dv$ is finite, one
can invoke the central limit theorem to assert that
${\mathcal P}(k,u)$ has a Gaussian distribution, i.e.,
\begin{equation}
{\mathcal P}(k,u)\approx \sqrt{\frac{k}{2\pi 
\sigma^2}}\,\exp\left[-\frac{ku^2}{2\sigma^2}\right].
\label{clt1}
\end{equation}
Substituting this result in Eq. (\ref{lsd2}) and replacing the resulting sum by an integral
in the small $s$ limit, we find that an appropriate scaling limit
of $Y(s,v)$ exists when $s\to 0$, $v\to 0$ keeping the ratio $v/\sqrt{s}$ fixed.
In this scaling limit, we get 
\begin{equation}
Y(s,v) \approx \frac{1}{2}\int_{s}^{\infty} \frac{dy}{y}\, e^{-y}\, {\rm 
erfc}\left(\frac{v\sqrt{y}}{\sigma \sqrt{2s}}\right)
\label{Ysvscaling}
\end{equation}
The integral on the r.h.s of Eq. (\ref{Ysvscaling}) can be done explicitly and 
one gets to leading order in small $s$
\begin{equation}
Y(s,v) \approx -\frac{1}{2}({\bf C}+\ln(s))-\ln\left[w+\sqrt{w^2+1}\right]
\label{Ysv2}
\end{equation}
where ${\bf C}= 0.57721566\ldots$ is the Euler's constant and $w= \frac{v}{\sigma \sqrt{2s}}$
is the scaling variable. Finally we substitute this expression for $Y(s,v)$  
and the limiting Gaussian form of ${\mathcal P}(n,v)$ from Eq. (\ref{clt1})
onto the r.h.s of Eq. (\ref{lsd1}), perform the subsequent integration and finally
arrive at the following expression in the scaling limit $s\to 0$, $n\to \infty$ but keeping
the product $ns$ fixed
\begin{equation}
\sum_N C_n(n) e^{-sN} \approx \frac{1}{\sqrt{n}}\, G(n\,s)
\label{lsds1}
\end{equation}
where the scaling function $G(x)$ is given by
\begin{equation}
G(x)= \frac{2 b}{\sqrt{\pi}}\, \frac{e^{-x}}{\sqrt{x}}\,\int_0^{\infty} dy\, e^{-y^2}\, 
\sqrt{1+ \frac{y^2}{x}}
\label{lsdsf1}
\end{equation} 
and
\begin{equation}
 b= e^{{\bf C}/2}=1.33456\ldots \, .
 \label{eq:defofb}
\end{equation} 
 We notice one remarkable fact: even though we
needed to have $\sigma^2$ finite in arriving at the result in Eq. (\ref{lsdsf1}),
$\sigma^2$ has dropped out of the final expression in Eq. (\ref{lsdsf1}). Thus for
any finite $\sigma^2$, the scaling function $G(x)$ is universal
and is actually independent of the value of $\sigma^2$.

From the expression in Eq. (\ref{lsdsf1}) it is easy to compute the moments
of the leader size distribution for large $N$. For example, to compute the $k$-th moment 
${\langle n^k\rangle}_N$ for 
large $N$, we multiply both sides of Eq. (\ref{lsdsf1}) by $n^k$ and sum over all $n$. This 
gives,
\begin{equation}
\sum_N {\langle n^k\rangle}_N \, e^{-sN} \approx \frac{2 b}{\sqrt{\pi}\sqrt{s}}\, \sum_n 
e^{-sn}\,n^{k-1} \int_0^{\infty} dy \, e^{-y^2}\, \sqrt{1+y^2/{ns}}.
\label{moment1}
\end{equation}
We next make a change of variable $y= \sqrt{sn}\, u$ in the integral and carry out
the sum over $n$. In the scaling limit, this sum can be replaced by an integral
which can be explicitly performed and we get for small $s$
\begin{equation}
\sum_N {\langle n^k\rangle}_N \, e^{-sN} \approx  \frac{2 b \Gamma(k+1/2)}{\sqrt{\pi}
s^{k+1/2}}\, \int_0^{\infty} \frac{du}{(1+u^2)^k}= b \frac{\Gamma(k-1/2)\Gamma(k+1/2)}{
\Gamma(k)}\,s^{-(k+1/2)}.
\label{moment2}
\end{equation}
Note that this expression is strictly valid for $k\ge 1$. For $k=0$, one can show
independently that $\sum_N C_N(n) e^{-sN}\to 1/s$ as $s\to 0$ as expected since
the PDF $C_N(n)$ is normalized to unity. Inverting the Laplace transform in Eq. 
(\ref{moment2}),it then follows that for large $N$, the $k$-th moment of the leader size
behaves for $k\ge 1$,
\begin{equation}
{\langle n^k\rangle}_N \approx A_k\, N^{k-1/2}; \quad\quad {\rm where}\quad \quad A_k= 
\frac{b\, 
\Gamma(k-1/2)}{\Gamma(k)}.
\label{moment3}
\end{equation}
In particular the average leader size $(k=1)$ behaves for large $N$ as
\begin{equation}
{\langle n\rangle}_N \approx b\,\sqrt{\pi}\, N^{1/2} = (2.63533\ldots)\, N^{1/2}
\label{avls}
\end{equation}

From the expression for the moments in Eq. (\ref{moment2}), it follows that
for large $N$, the leader size distribution $C_N(n)$ has the following scaling form
\begin{equation}
C_N(n) \approx \frac{1}{N^{3/2}}\, W\left(\frac{n}{N}\right) 
\label{scale1}
\end{equation}
where the scaling function $W(x)$, for $0\le x\le 1$ is such that, for all $k\ge 1$,
\begin{equation}
\int_0^1 W(x)\, x^k\, dx= b\, \frac{\Gamma(k-1/2)}{\Gamma(k)}
\label{scale2}
\end{equation}
Finally, one can analytically continue the Eq. (\ref{scale2}) for noninteger $k$
and then invert the equation to obtain the following explicit expression of $W(x)$
valid for $0\ll x \le 1$
\begin{equation}
W(x)= \frac{b}{\sqrt{\pi}}\, x^{-3/2} (1-x)^{-1/2}
\label{scale3}
\end{equation}
It is easy to verify that for all $k\ge 1$, the above expression for $W(x)$
does indeed satisfy Eq. (\ref{scale2}). 
Note that the scaling in Eq. (\ref{scale1}) is valid
only for $1\ll n \sim N$. In particular it does not hold when $n\sim O(1)$. This 
is also evident from the explicit form of the scaling function $W(x)$ in Eq. (\ref{scale3})
which diverges as $x^{-3/2}$ as $x\to 0$, making it non-normalizable. Indeed, for small
$n$, one has to keep the full expression of the generating function to recover
the correct normalization. 

Thus the main result of this subsection is the universal scaling form of the
leader size distribution in Eqs. (\ref{scale1}), (\ref{scale2}) and (\ref{scale3}) valid for 
any 
continuous and symmetric $\phi(v)$ with a
finite 
$\sigma^2$ and remarkably it is independent of the actual value of the $\sigma^2$.

\subsection{The Limiting Leader Velocity Distribution}

The marginal velocity PDF of the leader $D_N(v)$ can be obtained
by summing Eq. (\ref{lpdf3}) over $n$. This gives
\begin{equation}
\sum_{N=1}^{\infty} D_N(v)\, z^N= \frac{1}{1-z}\,\rho_z(v)\, \exp\left[-\int_v^{\infty} 
\rho_z(u) 
du\right].
\label{ldv1}
\end{equation}
A somewhat more convenient quantity is the cumulative velocity distribution of the
leader, $F_N(v)= \int_{-\infty}^v D_N(v')\, dv'$. One can easily derive its generating 
function by integrating Eq. (\ref{ldv1})
\begin{equation}
\sum_{N=1}^{\infty} F_N(v)\, z^N= \frac{\exp\left[-\int_v^{\infty}
\rho_z(u)
du\right]-(1-z)}{1-z}.
\label{cdv1}
\end{equation}
Note that as $v\to \infty$, the r.h.s of Eq. (\ref{cdv1}) becomes $z/(1-z)$.
This is consistent with the l.h.s of Eq. (\ref{cdv1}) since
$F_N(v)\to 1$ as $v\to \infty$ and thus the generating function in that limit
is precisely $z/(1-z)$. 

The result for the cumulative velocity distribution of the leader in
Eq. (\ref{cdv1}) is exact for all $N$. We next address the 
question of the limiting leader velocity distribution
as $N\to \infty$. For large $N$, one needs to investigate the r.h.s
of Eq. (\ref{cdv1}) in the limit $z\to 1$. In that limit, the r.h.s
scales as $\exp\left[-\int_v^{\infty}
\rho_1(u)
du\right]/(1-z)$, provided the numerator is finite. In such cases,
it follows from Eq. (\ref{cdv1}) that $F_N(v)$ tends to a limiting $N$-independent 
distribution $F_{\infty}(v)$ given by
\begin{equation}
F_{\infty}(v)=\exp\left[-\int_v^{\infty} \rho_1(u)\, du\right]= \exp\left[-\sum_{n=1}^{\infty} 
\frac{1}{n} \int_v^{\infty} {\mathcal P}(n,u)\, du\right].
\label{cdv2}
\end{equation} 
Note that for some $\phi(v)$ this limiting distribution may not exist.
In fact, we will see later that when $\phi(v)$ has the Cauchy distribution,
which is symmetric and continuous, there is no $N$-independent limiting distribution.
It is evident from Eq. (\ref{cdv2}) that, unlike the limiting leader
size distribution, the limiting leader velocity distribution, whenever it exists, is highly 
nonuniversal
and depends explicitly on the initial velocity distribution $\phi(v)$.

\begin{figure}
\includegraphics[width=3.375in]{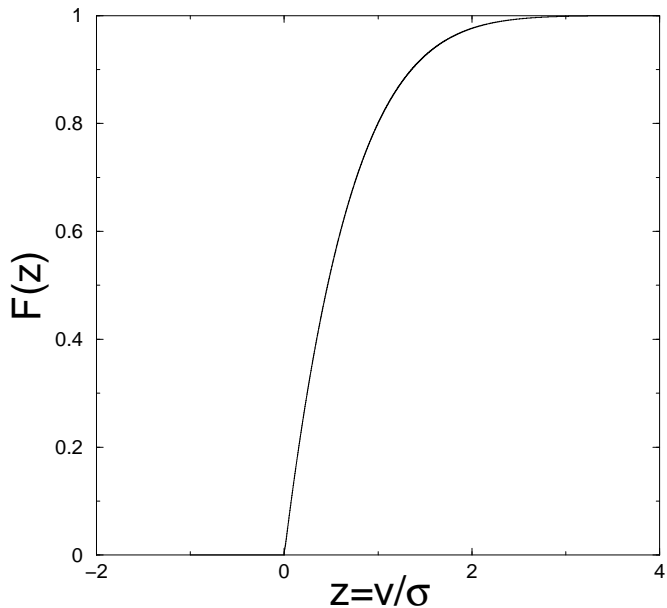}
\caption{\label{leadvelplot} (Color online).
The limiting cumulative velocity distribution $F_{\infty}(v)= F(z=v/\sigma)$
plotted against $z$, for the Gaussian distribution $\phi(v)=\exp(-v^2/{2\sigma^2})/\sqrt{2
\pi\sigma^2}$.}
\end{figure}

To see how this limiting distribution in Eq. (\ref{cdv2}) looks like, we 
first consider the Gaussian distribution, $\phi(v)= \exp(-v^2/{2\sigma^2})/\sqrt{2 
\pi\sigma^2}$. In this case, ${\mathcal P}(n,u)=\sqrt{\frac{n}{2\pi
\sigma^2}}\,\exp\left[-\frac{nu^2}{2\sigma^2}\right]$ exactly for all $u$.
Substituting on the r.h.s of Eq. (\ref{cdv2}) and performing the integral
one can express the limiting distribution $F_{\infty}(v)$ as a
function of the dimensionless variable $v/\sigma$,
\begin{equation}
F_{\infty}(v)= F(v/\sigma);\quad\quad {\rm where}\quad\quad F(z)= 
\exp\left[-\frac{1}{2}\sum_{n=1}^{\infty} \frac{1}{n}\, {\rm 
erfc}\left(\sqrt{\frac{n}{2}}\,z\right)\right].
\label{cdv.gauss1}
\end{equation}
While we were unable to perform the sum exactly in Eq. (\ref{cdv.gauss1}), the
function $F(z)$ can be easily plotted using Mathematica and is shown in Fig. 
(\ref{leadvelplot}). It is easy to show from Eq. (\ref{cdv.gauss1}) that
the function $F(z)$ has finite nonzero support only for positive $z$. It is
identically zero for all $z\le 0$. Furthermore, the asymptotics of $F(z)$
as $z\to 0$ and $z\to \infty$ can also be derived. We find
\begin{eqnarray}
F(z) & \approx & \sqrt{2}\, z \quad\quad\quad\quad\quad\quad\quad {\rm as}\quad z\to 0 
\label{z0} 
\\
&\approx & 1- \frac{1}{\sqrt{2\pi} \,z}\,e^{-z^2/2} \quad\,\,\, {\rm as}\quad z\to \infty.
\label{zinfty}
\end{eqnarray}
The asymptotic behavior as $z\to \infty$ is easy to derive, as in this limit the dominant
contribution comes from the $n=1$ term of the sum in Eq. (\ref{cdv.gauss1}). In contrast,
the other limit $z\to 0$ in Eq. (\ref{z0}) is more tricky to derive. We provide a derivation
in Appendix-B.

Let us make a couple of interesting observations on the general formula for the limiting 
velocity distribution in Eq. (\ref{cdv2}) when it exists.

\vskip 0.2cm

\noindent (i) For any continuous and symmetric initial velocity distribution
$\phi(v)$ for which the limiting velocity distribution $F_{\infty}(v)$ exists,
$F_{\infty}(v)=0$ for any $v\le 0$. This follows from the fact for $v=0$, by symmetry,
$\int_0^{\infty} {\mathcal P}(n,u)\,du=1/2$. Thus, the sum in Eq. (\ref{cdv2}) diverges
for $v=0$ and $F_{\infty}(0)=0$. Since $F_{\infty}(v)$ is a cumulative distribution, it
is a nondecreasing function of $v$. Hence it follows that if $F_{\infty}(v=0)=0$
then $F_{\infty}(v)=0$ for all $v\le 0$. The implication of this result
is rather interesting. It implies that even though initial 
velocity distribution may be symmetric in $v$ and there may be a lot of particles
initially with a negative velocity, the eventual velocity of the leader, i.e.,
the rightmost cluster is always positive.  

\vskip 0.2cm

\noindent (ii) Note that the velocity of the leader is also the maximum of the final
velocities. Now if the collisions were elastic, the particles 
would have
merely interchanged the velocities in each collision (this is called
the Jepsen gas), and in the final
state the velocity of the leader would have been the maximum of all the
initial velocities~\cite{BM1}. Therefore, in this case the distribution of the leader
velocity would have been given by the usual extreme value statistics of
uncorrelated random variables~\cite{EVT}. The velocity of
the final leader of elastic Jepsen gas scales as $u\approx a_N\, w+ b_N$ for
large $N$, and the limiting distribution of the scaled velocity $w$ has
only three possible forms depending on the tail of $\phi(v)$~\cite{BM1}.  However, the
scaling parameters $a_N$ and $b_N$ depends on $N$ as well as on the
complete form of $\phi(v)$. For example, when $\phi(v)$ is Gaussian,
$a_N=(2\ln N)^{-1/2}$ and $b_N=(2\ln N)^{1/2} - (2\ln N)^{-1/2} (\ln\ln N
+\ln 4\pi)/2$. The limiting velocity distribution is Gumbel in this case.
Thus, unlike the elastic gas where the leader velocity 
increases with $N$, it follows from Eq. (\ref{cdv2}) (when the limiting distribution exists) 
that in the sticky
gas, the leader velocity remains of $\Or(1)$ in the large $N$ limit.
This is due to the strong correlations generated between the velocities
of different clusters in the fan state. 
Thus Eq. (\ref{cdv2}) provides an exactly solvable case of extreme value 
statistics of correlated random variables.

\subsection{ Exact Leader Statistics for the Cauchy Distribution}

In this subsection, we consider the special case of the symmetric Cauchy distribution for the
initial velocity, 
\begin{equation}
\phi(v) = \frac{1}{\pi}\, \frac{a}{v^2+a^2}.
\label{cauchy1}
\end{equation}
For this distribution, the mean is zero, but the second moment $\sigma^2$ diverges.
So, the results of the previous two subsections, where it was assumed $\sigma^2$ is finite,
do not hold. However, the principal result in Eq. (\ref{lpdf3}) is still valid
and we show here that the Cauchy distribution preents an excatly solvable case
in the sense that the r.h.s of Eq. (\ref{lpdf3}) can be exactly evaluated in closed form.

The first simplification occurs when one calculates ${\mathcal P}(n,u)$ defined in
Eq. (\ref{def:mathcalP}). Since the Cauchy distribution is stable, it turns out
(as can be easily proved) that 
\begin{equation}
{\mathcal P}(n,u)= \phi(u) =\frac{1}{\pi}\, \frac{a}{u^2+a^2}.
\label{cauchy2}
\end{equation}
Thus ${\mathcal P}(n,u)$ is completely independent of $n$. This particular
feature of the Cauchy distribution was also used recently to obtain
excat results for the statistics of records in a sequence of random walk in
presence of a drift~\cite{LW}.
Using this independence on $n$, it then follows from the definition
in Eq. (\ref{defrho}) that
\begin{equation}
\rho_z(u)= -\frac{1}{\pi}\, \frac{a}{u^2+a^2}\,  \ln(1-z).
\label{cauchy3}
\end{equation} 
One can then trivially do the integral
\begin{equation}
\int_v^{\infty} \rho_z(u)\, du = 
-\left[\frac{1}{2}-\frac{1}{\pi}\arctan(v/a)\right]\,\ln(1-z).
\label{cauchy4}
\end{equation}
Substituting these results in Eq. (\ref{lpdf3}) gives the following exact expression
\begin{equation}
\sum_{N=1}^{\infty} L_N(n,v)\, z^N= \frac{1}{\pi}\, 
\frac{a}{v^2+a^2}\,\frac{z^n}{n}\,(1-z)^{-\mu(v)}\quad\quad {\rm where}\quad 
\mu(v)=\frac{1}{2}+ \frac{1}{\pi}\arctan(v/a).
\label{cauchy5}
\end{equation} 
Comparing coefficient of $z^N$ gives us the exact joint distribution of size and the velocity 
of the leader
\begin{equation}
L_N(n,v)= \frac{1}{\pi}\,
\frac{a}{v^2+a^2}\, \frac{\Gamma(\mu(v)+N-n)}{n\,\Gamma(N-n+1)\,\Gamma(\mu(v))}.
\label{cauchy6}
\end{equation}

\vskip 0.2cm

\noindent {\bf Marginal size distribution:} To compute the marginal size distribution $C_N(n)=
\int_{\infty}^{\infty} L_N(n,v)\, dv$ of the leader, it is convenient to do the
integration over $v$ in Eq. (\ref{cauchy5}). This gives
the exact generating function
\begin{equation}
\sum_{N=1}^{\infty} C_N(n)\, z^N= -\frac{z^{n+1}}{n\,(1-z)\,\ln(1-z)}.
\label{csize1}
\end{equation}
From this generating function, it is easy to calculate the moments ${\langle n^k\rangle}_N$
of the leader size for large $N$ by investing the singularity of the r.h.s in Eq. 
(\ref{csize1})
near $z=1$. Skipping the details, we find that for large $N$ and for $k\ge 1$
\begin{equation}
{\langle n^k\rangle}_N \approx \frac{1}{k}\, \frac{N^k}{\ln(N)}.
\label{csize2}
\end{equation}
This result should be compared to the case in Eq. (\ref{moment3}) where $\sigma^2$ is finite.
In particular, the average leader size grows as ${\langle n\rangle}_N \approx N/{\ln N}$, much 
faster than the $N^{1/2}$ growth for the case where $\sigma^2$ is finite.

\vskip 0.2cm
\noindent {\bf Marginal velocity distribution:} Similarly, the marginal velocity PDF
of the leader $D_N(v)= \sum_n L_N(n,v)$ can be computed conveniently directly from Eq. 
(\ref{cauchy5}). We get
\begin{equation}
\sum_{N=1}^{\infty} D_N(v)\, z^N= \frac{1}{\pi}\,
\frac{a}{v^2+a^2}\, [-\ln(1-z)]\, (1-z)^{-\mu(v)},
\label{cvel1}
\end{equation}
from which one can compute the generating function for the cumulative velocity distribution
$F_N(v)= \int_{-\infty}^v D_N(v')\, dv'$. We get a nice compact expression
\begin{equation}
\sum_{N=1}^{\infty} F_N(v)\, z^N = (1-z)^{-\mu(v)}-1.
\label{cvel2}
\end{equation}
Comparing powers of $z^N$ gives a very simple but nontrivial distribution valid for all $N\ge 
1$
\begin{equation}
F_N(v)= \frac{\Gamma(\mu(v)+N)}{\Gamma(N+1)}; \quad\quad {\rm where}\quad \mu(v)= 
\frac{1}{2}+ \frac{1}{\pi}\arctan(v/a).
\label{cvel3}
\end{equation}
Note that when $v\to \infty$, $\mu(v)\to 1$ and when $v\to -\infty$, $\mu(v)\to 0$.
Thus, in the limit $v\to \infty$, $F_N(v)\to 1$ and when $v\to -\infty$, $F_N(v)\to 0$
as expected. Also, for $N=1$, we get $F_1(v)= \mu(v)$ which is just the cumulative 
Cauchy distribution. This is expected, because the velocity of a single particle remains 
unchanged as there is no collision.

We note from this exact velocity distribution in Eq. (\ref{cvel3})
that the distribution is $N$-dependent explicitly and does not have any nontrivial 
$N$-independent limiting 
form
as $N\to \infty$, unlike in the previous subsection for the Gaussian case. In fact, by 
ananlysing the singularity near $z=1$ in Eq. (\ref{cvel2}), we see that for any fixed $v$,
as $N\to \infty$, $F_N(v)$ decays with $N$ as a power law with an exponent that
depends continuously on $v$
\begin{equation}
F_N(v)\approx \frac{1}{\Gamma(\mu(v))}\, \frac{1}{N^{1-\mu(v)}}.
\label{cvel4}
\end{equation}
Since the exponent $1-\mu(v)= 1/2-\arctan(v/a)/\pi$ decreases monotonically
from $1$ (as $v\to -\infty$) to zero (as $v\to \infty$), it follows that
the distribution for smaller values of $v$ tends to zero faster than
the distribution for higher $v$.
Thus the marginal leader velocity distribution in the Cauchy case is very different
from that of its Gaussian counterpart.

\section{Distribution of the Total Energy in the Fan State}

In this section we compute the distribution of the total energy
of the clusters in the fan state. For simplicity, we restrict ourselves
to the initial Gaussian distribution of the velocities, 
$\phi(v)=  \exp(-v^2/{2})/\sqrt{2
\pi}$ where we have also set $\sigma^2=1$ without any loss of generality.
Thus, initially, 
the total (kinetic) energy of the system, $E_0=\sum_{i=1}^N v_i^2/2$, is
distributed according to the PDF
\begin{equation}
P_N(E_0)
=\frac{E_0^{N/2-1}\,\rme^{-E_0}}{\Gamma(N/2)}.
\label{initial energy}
\end{equation}
Now, as the system evolves, it dissipates energy through collisions, and once
the fan  state is reached there is no more energy loss. It is natural
 to ask how the PDF of the energy evolves with time from its initial form in Eq. (\ref{initial 
energy}), and in particular how does this energy PDF look like 
in the final fan state beyond which there is no more dissipation.

Let $Q_N(E)$ denote the PDF of the total energy in the fan state, i.e.,
\begin{equation}
Q_N(E)=\left\langle\delta\left(E-\frac{1}{2}\sum_{i=1}^k n_i\, u_i^2\right)
\right\rangle,
\label{final energy}
\end{equation}
where the angle brackets $\langle \dotsb\rangle$ denote the average over
the random variables $k, \{n_i\}$ and $\{u_i\}$, whose joint PDF is given
by \eref{jointpdf}. It is convenient to consider the Laplace transform
of the energy   
in Eq. (\ref{final energy})
\begin{equation}
\int_0^\infty Q_N(E)\, \rme^{-\alpha E}\, \rmd E =\left\langle 
\exp\left[-\frac{\alpha}{2}\sum_{i=1}^k n_i\, u_i^2\right]\right\rangle.
\label{lt1}
\end{equation}
To evaulate this average using the measure in Eq. (\ref{jointpdf}), we need
to first evaluate ${\mathcal P}(n,u)$ in Eq. (\ref{def:mathcalP}). For
Gaussian $\phi(v)$, this is simple: ${\mathcal P}(n,u)=\sqrt{\frac{n}{2\pi}}\,
\exp\left[-\frac{n\,u^2}{2}\right]$ exactly for all $n$ and $u$.
We substitute this in the measure in Eq. (\ref{jointpdf}) and note
that calculating the average in Eq. (\ref{lt1}) just amounts to
renormalizing the effective ${\mathcal P}(n,u)\to \sqrt{\frac{n}{2\pi}}\,
\exp\left[-(1+\alpha)\frac{n\,u^2}{2}\right]$. With this input, one can then carry out
exactly the same procedure as in subsection III A and one arrives at the final result 
\begin{equation}
\int_0^\infty Q_N(E)\, \rme^{-\alpha E}\, \rmd E =\sum_{k=1}^N
(1+\alpha)^{-k/2}\; 
\Biggl[\frac{S_1(N,k)}{N!}\Biggr],
\label{laplace transform E}
\end{equation}
where we recall $S_1(N,k)/N!$ from \eref{pk}.  The inverse Laplace transform
yields
\begin{equation}
Q_N(E)=\sum_{k=1}^N\Biggl[\frac{E^{k/2-1}\,\rme^{-E}}{\Gamma(k/2)}\Biggr]
\cdot\Biggl[\frac{S_1(N,k)}{N!}\Biggr].
\label{exact final energy}
\end{equation}
Since in the final
state the number of clusters $k$ is distributed according to
$p_N(k)=S_1(N,k)/N!$, it is interesting to note that, for a given number
$k$ of final clusters, the distribution of the total energy is identical to
that of $k$ initial particles.

\Eref{exact final energy} provides an exact expression for $Q_N(E)$, for
all values of $E$ and $N$.  However, it is always desirable to have an
closed-form expression, even though its validity would require taking the
limit of large $N$. To achieve this goal, we first express \eref{laplace
transform E} in the following form by using the generating function given
in \eref{stirling gen 2}:
\begin{equation}
\int_0^\infty Q_N(E)\, \rme^{-\alpha E}\, \rmd E = 
\frac{1}{N!}\, \frac{\Gamma\Bigl(N+1/\sqrt{1+\alpha}\Bigr)}
{\Gamma\Bigl(1/\sqrt{1+\alpha}\Bigr)}.
\label{laplace transform E (2)}
\end{equation}

At this point, we make a short detour to compute the mean and variance of
the final total energy, as they are readily available from the first and
second derivatives of \eref{laplace transform E (2)} with respect to
$\alpha$ at $\alpha=0$. We find the mean as
\begin{equation}
\langle E \rangle = \frac{1}{2} H_N
\label{mean energy}
\end{equation}
where $H_{N}=\sum_{k=1}^N k^{-1}$ is the harmonic number, and
 asymptotically, $H_N\sim\ln N+{\bf C}$ where ${\bf C}$ is the Euler's constant. The variance 
is 
\begin{equation}
\langle E^2 \rangle - \langle E \rangle^2=\frac{3}{4} H_N
-\frac{1}{4} H_{N,2}
\label{variance of energy}
\end{equation}
where $H_{N,r}=\sum_{k=1}^N k^{-r}$ is generalized harmonic number. 
Note that $H_N=H_{N,1}$.
The limiting value of $H_{N,r}$
as $N\to \infty$ is the Riemann zeta function $H_{\infty,r}=\zeta(r)$, and
$\zeta(2)=\pi^2/6$.

We  now return to \eref{laplace transform E (2)}, and proceed along the
main course to obtain a closed-form expression of $Q_N(E)$ for large $N$.
The inverse Laplace transform of \eref{laplace transform E (2)} is formally
given by the Bromwich integral
\begin{equation}
Q_N(E)=\frac{1}{2\pi i} \int_{c-i\infty}^{c+i\infty}
\rme^{S(\alpha)}\,\rmd\alpha,
\label{Bromwich integral}
\end{equation}
where
\begin{align}
S(\alpha)=&\ln \Gamma\Bigl(N+1/\sqrt{1+\alpha}\, \Bigr) 
-\ln \bigl(N!\bigr)
\notag\\
&-\ln\Gamma\Bigl(1/\sqrt{1+\alpha}\, \Bigr)
  +\alpha E.
\label{S alpha}
\end{align}
%

Now a saddle point approximation of the integral in \eref{Bromwich
integral} yields
 \begin{equation}
 Q_N(E)\approx \rme^{S(\alpha^*)}\Big/\sqrt{2\pi \abs{S''(\alpha^*)}}\; ,
 \label{approximate final  energy}
 \end{equation}
 in which $\alpha^*$ is determined implicitly in terms of $E$ and $N$ by
 the saddle point condition $S'(\alpha^*)=0$.
Enforcing this condition to \eref{S alpha} gives the saddle point equation
for $\alpha^*$ as
 \begin{equation}
 E=\frac{\beta^3}{2} \Bigl[
   \psi(N+\beta)-\psi(\beta)\Bigr],
 \label{saddle condition}
\end{equation}
where $\beta=1/\sqrt{1+\alpha^*}$ and $\psi(x)=(\rmd/\rmd x) \ln
 \Gamma(x)=\Gamma'(x)/\Gamma(x)$ is the digamma Function. As
 $\psi(N+\beta)\sim\ln N$, \eref{saddle condition} suggests that when both
 $E$ and $N$ are large, the natural scaling variable is $z=2E/\ln N$.  In
 terms of this scaling variable, for large $N$, from \eref{saddle
 condition} we find
 \begin{equation}
 \beta=z^{1/3} +\frac{z^{1/3}}{3\ln N} \psi(z^{1/3}) 
 + \Or \left(\frac{1}{[\ln N]^2}\right).
 \label{beta}
 \end{equation}

Now, employing Stirling's approximation (for large $N$) in \eref{S alpha}
and using \eref{beta}, in terms of $z$ we have
 \begin{equation}
 S(\alpha^*)=\left[\frac{3}{2}z^{1/3}-\frac{z}{2}-1\right]\ln N
 -\ln\Gamma(z^{1/3})
 + \Or \left(\frac{1}{\ln N}\right).
\label{S final}
 \end{equation}
 Similarly, we find
 \begin{equation}
 S''(\alpha^*)\sim  \frac{3}{4} z^{5/3}\, \ln N.
\label{S'' final}
 \end{equation}

\begin{figure}
  \includegraphics[width=3.375in]{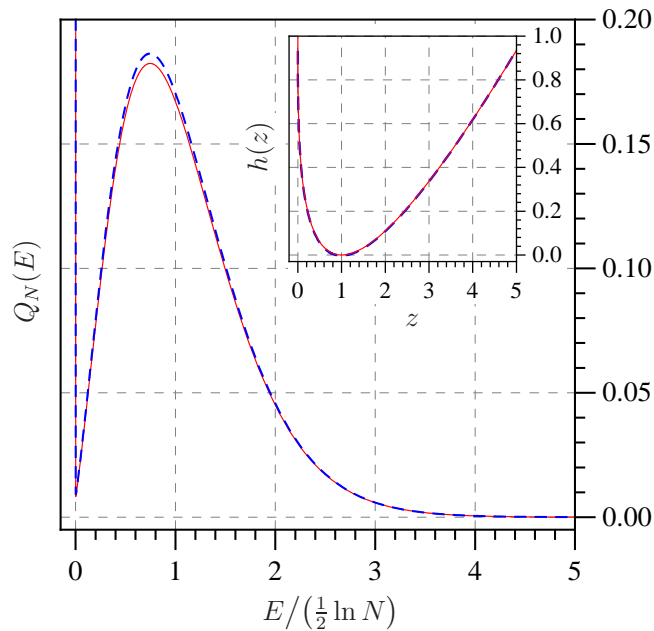}
\caption{\label{energyplot} (Color online). Main: Distribution of the
total energy for $N=2000$.  The solid (red) line plots the saddle-point
approximation \eref{Q_N for large N}, and the dashed (blue) line plots the
exact form \eref{exact final energy}, evaluated using Mathematica. Inset:
The solid (red) line plots the large deviation function \eref{large
deviation function}, and the dashed (blue) line plots $-(\ln N)^{-1}\ln
Q_N\bigl(\frac{z}{2}\ln N\bigr) -(\ln N)^{-1}\ln \bigl[(\frac{3\pi}{2} \ln
N )^{1/2} z^{5/6}\,\Gamma( z^{1/3}) \bigr]$ as a function of $z$, using the
exact form of $Q_N(E)$ given in \eref{exact final energy}.}
\end{figure}

Therefore, substituting $S$ and $S''$ form \eref{S final} and \eref{S''
 final} respectively, in \eref{approximate final energy}, finally we obtain
\begin{equation}
Q_N\biggl(\frac{z}{2}\ln N\biggr)\approx \left(\frac{3\pi}{2} \ln N
\right)^{-1/2} \frac{\rme^{-h(z)\,\ln N}}{z^{5/6}\,\Gamma(
z^{1/3})},
\label{Q_N for large N}
\end{equation}
where the large deviation function $h(z)$ is given by
\begin{equation}
h(z)= 1+\frac{z}{2}-\frac{3}{2}z^{1/3}.
\label{large deviation function}
\end{equation}
The saddle-point approximation \eref{Q_N for large N} is indeed a very good
for large $N$, as we demonstrate in \fref{energyplot} by comparing with
the exact result \eref{exact final energy} for $N=2000$.

Near $z=1$ we have $h(1+x)= x^2/6 +\Or (x^3)$. Therefore, from \eref{Q_N
for large N} we find that $Q_N(E)$ has a Gaussian form close to the mean
$\langle E\rangle \sim (1/2)\ln N$, with a variance $\langle E^2 \rangle -
\langle E \rangle^2\sim (3/4)\ln N$; ---we note that by taking the large
$N$ limit in \eref{mean energy} and \eref{variance of energy} respectively,
we independently recover these mean and variance.  However, far away from
the mean, the PDF is quite asymmetric as clearly seen in \fref{energyplot}.

\section{Summary and Conclusion}

In summary, we have studied analytically various statistical properties in
the final fan state of a one dimensional sticky gas of $N$ particles. The particles are
initially distributed randomly in space, each with unit mass and with an initial velocity 
drawn independently from an identical distribution $\phi(v)$.
Each particle moves ballistically and when two particles collide, they form
a single cluster with the total mass and total momentum conserved in the collision
process. In the long time limit, the system reaches a fan state where the
system consists of a finite number of clusters with their velocities 
increasing from left to right and there is no further collision.
We have shown that the sizes and the number of clusters are distributed
universally in the fan state, independent of $\phi(v)$ and this distribution
is identical to that of the cycles lengths in random permutation problem.
We have also computed exactly the distribution of the size
and the velocity of the rightmost cluster (leader) moving with the
largest velocity. Furthermore, the distribution of the total energy
in the fan state is also computed exactly for Gaussian $\phi(v)$.  
Our results provide an exactly solvable case of a many-body system (of {\em finite} size $N$)
with dissipative dynamics and in particular, brings out the universal
features of the final stationary state in an explicit way.

There are interesting open issues for further research. Here we have focused
solely on the final fan state. It would be interesting to study the
dynamics, i.e., the approach to this fan state and investigate
to what extent the universal features exist away from the fan state.

There are also immediate generalizations of the 
problem studied here that are open for future research.
For example, it would be interesting to know to what extent the
universal features in the fan state remain valid for an
inhomogeneous sticky gas. The inhomogeneity can arise
either from unequal initial masses 
or nonidentical velocity distributions in the initial condition.
For an infinite system, i.e., with a finite {\em density} of initial particles, 
the ballistic 
aggregation model with inhomogeneous
initial condition has been studied~\cite{FJM}. It would be interesting to
extend this study to the case of a finite number $N$ of particles. 

Finally, it would be interesting to study this finite system of 
$N$ particles when the collisions are inelastic, but not necessarily sticky, i.e.,
with a nonzero coefficient of restitution. 
This problem has been studied for an infinite system
and many asymptotic properties such as the energy decay and the velocity distribution
were found to be similar to the sticky gas limit~\cite{bennaim1,shinde}.
It would be interesting to see to what extent the universal features
in the stationary state for finite $N$ are retained when one introduces
a nonzero coefficient of restitution.

\vskip 0.5cm

\noindent {\bf Acknowledgements:} We thank D. Dhar for many stimulating discussions
and for his early participation in this work. We acknowledge
useful discussions with A. Comtet, P.~L. Krapivsky and Y. Peres.
The support from grant no. 3404-2 of ``Indo-French Center
for the Promotion of Advanced Research (IFCPAR/CEFIPRA)"
is also gratefully acknowledged.

\appendix

\section{Proof of Equation~(\ref{Prclustersizes})}

 In this Appendix, we derive the distribution of the cluster 
 sizes (regardless of their positions)
 starting from  joint probability distribution  of the
  cluster sizes and velocities  given in  Eq.~(\ref{jointpdf}).
 To  calculate $\text{Pr}\{c_1,c_2,\dotsc,c_N \}$
 from the joint-PDF $p_N(k;\{n_i,u_i\})$, we must: 
 (i) integrate out the velocities,  
 (ii) sum over all configurations with the same  cluster sizes 
  (but placed in different orders)  
 (iii) change the variables from the list  $(n_1,,\dotsc,n_k)$
  that  represents  the sizes  of the $k$  consecutive clusters 
  to the set $\{c_1,\dotsc,c_N\}$
  that  encodes  the number of clusters of a given size. We thus have 
\begin{equation}
\text{Pr}\{c_1,c_2,\dotsc,c_N \}= 
  \sum_{ \substack{ \hbox{Permutations of} \\  \hbox{unequal size  clusters} } } 
 p_N(n_1,\ldots,n_k) \, , 
\end{equation}
 where the sum is restricted to   permutations that exchange clusters having different
 sizes i.e.   permutations   of the list $(n_1,\ldots,n_k)$
that exchange  $n_i$'s  having  different values.  Using  Eq.~(\ref{Nbcluster1})
  and the fact  that  $\prod_{i=1}^k n_k =  \prod_{j=1}^N j^{c_j}$,
  we find that  $p_N(n_1,\ldots,n_k)$   is given by
\begin{equation}
p_N(n_1,\ldots,n_k) = \frac{\delta\left(N-\sum_{j=1}^N j c_j  \right)}
 {\prod_{j=1}^N j^{c_j}} \, 
  \int   \prod_{\ell=1}^{k} du_\ell \,  {\mathcal P}(n_\ell, u_\ell) 
  \prod_{\ell=1}^{k-1}\theta(u_{\ell+1}-u_\ell) \, .
\label{eq:App2}
\end{equation}
  From the definition of  ${\mathcal P}(n_\ell, u_\ell)$  given in 
 Eq.~(\ref{def:mathcalP}), 
 the integral  on the r.h.s of Eq.~(\ref{eq:App2}) is found to be 
\begin{eqnarray}
  \int \prod_{i=1}^N dv_i \, \phi(v_i) \,\,
   \prod_{\ell=1}^{k-1}   \theta\left( 
    \frac{1}{n_{\ell+1}} {\sum_{j=N_{\ell}+1}^{N_{\ell+1}}  v_j} -
     \frac{1}{n_\ell}  {\sum_{j=N_{\ell-1}+1}^{N_\ell}  v_j} 
 \right)
\end{eqnarray}
 where  $N_{\ell}= n_1 + \ldots + n_{\ell}$ for $\ell=1,\ldots,k$.
 The proof of Eq.~(\ref{Prclustersizes}) hence reduces to showing  the
 following identity:
   \begin{eqnarray}
   \sum_{ \substack{ \hbox{Permutations of} \\  \hbox{unequal clusters} } }  
  \int \prod_{i=1}^N dv_i \, \phi(v_i) \,\,
   \prod_{\ell=1}^{k-1}   \theta\left( 
    \frac{1}{n_{\ell+1}} {\sum_{j=N_{\ell}+1}^{N_{\ell+1}}  v_j} -
     \frac{1}{n_\ell}  {\sum_{j=N_{\ell-1}+1}^{N_\ell}  v_j} 
 \right)  =  \frac{1}{c_1!\, c_2! \ldots c_N!} \,, 
\label{Idtobeproved} 
\end{eqnarray}
where we recall that $c_1$ is equal to the total number of 1's in the list
 $(n_1,\ldots,n_k)$,  $c_2$ is equal to the total number of 2's in this list etc...
 This identity is proved by the following argument: consider a permutation
 that exchanges two clusters of the same size $p$;  then,  by relabeling the $v_i$
(which are dummy variables) we observe that the integral on the left hand side (l.h.s) of
 Eq.~(\ref{Idtobeproved}) is invariant. There are $c_p!$ possible  permutations 
of  the  $c_p$ clusters of size $p$. Thus we have formally
 \begin{eqnarray}
      c_1!\, c_2! \ldots c_N!
   \sum_{ \substack{ \hbox{Permutations of} \\  \hbox{unequal clusters} } }  
 =  \sum_{ \hbox{All Permutations} } 
  \, .
 \label{2sums}
\end{eqnarray}
  If  we have $k$ distinct numbers $u_1 \ldots u_k$ there is only one
 permutation that orders them   in increasing  order. Therefore, we have 
 \begin{eqnarray}
     \sum_{\sigma \in \Sigma_k} 
   \prod_{\ell=1}^{k-1}\theta(u_{\sigma(\ell+1)}-u_{\sigma(\ell)} ) = 1 \, , 
\end{eqnarray}
where  $\Sigma_k$ is the symmetric group of $k$ elements.
After  substituting  this identity in Eq.~(\ref{2sums}) and  using the fact that
 the PDF  $\phi(v_i)$ is normalized,    Eq.~(\ref{Idtobeproved}) is  finally proved.
We note that the  method of proof used here  is closely related to the  derivation by F.~Spitzer
 of  a generalization of the Sparre-Andersen theorem~\cite{spitzer}.

\section{Limiting Behavior of $F(z)$ as $z\to 0$ in Eq. (\ref{cdv.gauss1})}

Let us write $F(z)$ in Eq. (\ref{cdv.gauss1}) as $F(z)=\exp[-S(z)]$ where
the sum
\begin{equation}
S(z)= \sum_{n=1}^{\infty}\frac{1}{2n}\, {\rm
erfc}\left(\sqrt{\frac{n}{2}}\,z\right).
\label{sum1}
\end{equation}
We want to compute the behavior of $S(z)$ as $z\to 0$. We first note the identity
\begin{equation}
\sum_{n=1}^{\infty} \frac{1}{2n}\, \exp(-n\, z^2/2)= 
-\frac{1}{2}\,\ln\left[1-e^{-z^2/2}\right].
\label{sum2}
\end{equation}
Subtracting Eq. (\ref{sum2}) from Eq. (\ref{sum1}) gives
\begin{equation}
S(z) +\frac{1}{2}\,\ln\left[1-e^{-z^2/2}\right]= \sum_{n=1}^{\infty}\frac{1}{2n}\,
\left[{\rm erfc}\left(\sqrt{\frac{n}{2}}\,z\right)-e^{-n\,z^2/2}\right].
\label{sum3}
\end{equation}
Now, let us define a new variable $y= n\, z^2/2$. As $n$ changes by $1$, $y$ changes
by $\Delta y= z^2/2$. Now, in the limit $z\to 0$, this increment $\Delta y$ is very small.
Thus, one can replace the sum over $n$ on the r.h.s of Eq. (\ref{sum3}) by an integral
over $y$ to leading order for small $z$, giving
\begin{equation}
S(z) +\frac{1}{2}\,\ln\left[1-e^{-z^2/2}\right] \approx I=\frac{1}{2}\,\int_0^{\infty} 
\frac{dy}{y}\, \left[{\rm erfc}\left(\sqrt{y}\right)-e^{-y}\right].
\label{sum4}
\end{equation}
The integral $I$ on the r.h.s is just a constant and can be evaluated by parts
\begin{equation}
I= -\frac{1}{2}\left[\int_0^{\infty}\, e^{-y}\, \ln(y)\, dy 
-\frac{1}{\sqrt{\pi}}\int_0^{\infty} 
\frac{\ln(y)}{\sqrt{y}}\, e^{-y}\, dy \right].
\label{sum5}
\end{equation}
The two integrals on the r.h.s are elementary ones and can be easily evaluated
to give finally, $I= -\ln(2)$. Substituting this in Eq. (\ref{sum3}) and taking the $z\to 0$ 
limit gives
\begin{equation}
S(z) \to -\ln(\sqrt{2}\, z),
\label{sum6}
\end{equation}
implying as $z\to 0$ to leading order
\begin{equation}
F(z)= \exp[-S(z)]\to \sqrt{2}\, z.
\label{sum7}
\end{equation}


\begin{thebibliography}{99}


\bibitem{carnevale}{G. F. Carnevale, Y. Pomeau, and W. R. Young,
Phys. Rev. Lett. {\bf 64}, 2913 (1990)}.

\bibitem{granular gas} {\em Granular Gases}, edited by T. P\"oschel and
S. Luding (Springer, Berlin, 2001); {\em Granular Gas Dynamics}, edited by
T. P\"oschel and N. Brilliantov (Springer, Berlin, 2003).

\bibitem{Shandarin} S. F. Shandarin and Ya. B. Zeldovich,
  Rev. Mod. Phys. {\bf 61}, 185 (1989).

\bibitem{martin} P.~A.~Martin, J.~Piasecki, 
 J. Stat. Phys,  {\bf 76}, 447  (1994); J. Stat. Phys,  {\bf 84}, 837  (1996).

\bibitem{burgers} J. M. Burgers, {\em The Nonlinear Diffusion Equation}
  (Reidel, Dordrecht,1974).

\bibitem{kida} S. Kida, J. Fluid Mech. {\bf 93} part 2, 337 (1979).

\bibitem{frisch} U.~Frisch and J.~Bec,  {\em ``Burgulence''},
 in  {\em Les  Houches 2000: New Trends in Turbulence}, M. Lesieur
 ed., (Springer EDP-Sciences).

\bibitem{frachebourg} L. Frachebourg, 
 Phys. Rev. Lett. {\bf 82}, 1502 (1999).


\bibitem{fracheb2} L. Frachebourg, P.~A.~Martin and  J.~Piasecki,
Physica A {\bf 279}, 69 (2000). 


\bibitem{shida}  K. Shida and T.  Kawai,  
  Physica A {\bf 162},  145  (1989).

\bibitem{sibuya} M. Sibuya, T.  Kawai, and K. Shida,
Physica A {\bf 167}, 676 (1990).

\bibitem{hyuga} H.~Hyuga, T.~Kawai, K.~Shida and S.~Yamada,
Physica A {\bf 241}, 664 (1997).

 \bibitem{goncharov} V. Goncharov, {\em Sur la distribution des cycles 
 dans les permutations},  C.R. (Doklady) Acad. Sci. URSS (N.S) {\bf 35},
 267 (1942).



\bibitem{shepp} L.A. Shepp, S.P. Lloyd, Trans. Am. Math. Soc. {\bf 121},
  340 (1966).

\bibitem{spitzer} F. Spitzer,  Trans. Am. Math. Soc.  {\bf 82}, 323 (1956).


 \bibitem{steele} J. M. Steele, J. Comput. Appl. Math. {\bf 142}, 235
  (2002).



\bibitem{PaulK}  E. Trizac and  P.~L.~Krapivsky,  
Phys. Rev. Lett. 91, 218302 (2003).


\bibitem{feller} W. Feller, {\em An Introduction to Probability Theory and
  Its Applications} (Wiley, New York, 1971), Vol. II, 2nd ed.


\bibitem{Knuth} R. L. Graham,  D. E. Knuth 
  and O. Patashnik,  {\em Concrete Mathematics} 
  (Addison-Wesley,  2nd ed., 1994). 

\bibitem{riordan}  J.~Riordan, {\em Introduction to Combinatorial
 Analysis}  (Dover,  New York, 2002).

\bibitem{Finch} S. R. Finch, {\em Mathematical Constants}, (Cambridge
University Press, UK, 2003).


\bibitem{arratia}  R. Arratia, A. D. Barbour and  S. Tavare,  
Notices of the AMS,  {\bf 44}, 903 (1997). 

\bibitem{MZ} S.N. Majumdar and R.M. Ziff, Phys. Rev. Lett. {\bf 101}, 050601 (2008).

\bibitem{GL} C. Godreche and J.M. Luck, arXiv: 0809.3377.

\bibitem{LW} P. Le Doussal and K.J. Wiese, arXiv: 0808.3217.

\bibitem{BM1} I. Bena and S.N. Majumdar, Phys. Rev. E {\bf 75}, 051103 (2007);
S. Sabhapandit, I. Bena, and S.N. Majumdar, JSTAT  P05012 (2008).

\bibitem{EVT} R.A.~Fisher and L.H.C.~Tippet, Proc. Cambridge Philos. Soc.
  {\bf 24}, 180 (1928);  E.J.~Gumbel, \emph{Statistics of Extremes}
  (Columbia University Press, NY, 1958).

\bibitem{FJM} L.~Frachebourg, V.~Jacquemet, and P.~A.~Martin, J. Stat. Phys. {\bf 105}, 745 
(2001).

\bibitem{bennaim1} E. Ben-Naim, S.Y. Chen, G.D Doolen, and S. Redner,
  Phys. Rev. Lett. {\bf 83}, 4069 (1999).

\bibitem{shinde} M. Shinde, D. Das, and R. Rajesh, Phys. Rev. Lett. {\bf
  99}, 234505 (2007).



\end{thebibliography}
\end{document}